\documentclass[prd,preprint,showpacs,groupedaddress]{revtex4-1}
\usepackage{amsmath}
\usepackage{amsfonts}
\usepackage{amssymb}
\usepackage{geometry}
\usepackage{natbib}
\usepackage[english]{babel}
\usepackage{graphicx}
\usepackage{epstopdf}
\usepackage{subfigure}
\usepackage{caption}
\usepackage{multirow}
\usepackage{indentfirst}
\usepackage{mathrsfs}
\usepackage{booktabs}
\usepackage[colorlinks,linkcolor=blue,anchorcolor=blue,citecolor=blue]{hyperref}

\begin{document}
	
	\setlength{\parindent}{2em}
	\title{Double shadow of a 4D Einstein-Gauss-Bonnet black hole and the  connection between them with quasinormal modes}
	\author{Tian-Tian Liu} \author{He-Xu Zhang} \author{Yu-Hang Feng} \author{Jian-Bo Deng} \author{Xian-Ru Hu} \email[Xian-Ru Hu: ]{huxianru@lzu.edu.cn}
	\affiliation{Institute of Theoretical Physics $\&$ Research Center of Gravitation, Lanzhou University, Lanzhou 730000, China}
	
\begin{abstract}
	In this paper, we study the shadow of a 4D Einstein-Gauss-Bonnet black hole as photons couple to the Weyl tensor and find that the propagation of light depends on its polarization which leads to the existence of a double shadow. Then, we discuss the effect of the coupling parameter  $\lambda$, the polarization of light and the Gauss-Bonnet coupling constant  $\alpha$ on the shadow. Further we explore the influence of the Gauss-Bonnet coupling constant  $\alpha$ on the quasinormal modes (QNMs) of massless scalar field and investigate the connection between the real part of QNMs in the eikonal limit and the shadow radius of black holes. We find that in the eikonal limit the real part of QNMs is inversely proportional to the shadow radius under the case of the photons uncoupled to the Weyl tensor.
\end{abstract}

  %\keywords{Black holes, shadow, quasinormal mode}
  \maketitle
  
\section{Introduction}	
  Einstein’s general theory of relativity predicted the existence of black holes, which has fascinated physicists over decades. Thus it is very important to detect black hole parameters for studying features of the black hole. Investigations show that observing the shadow of black hole provides a tentative way to obtain information as regards black hole \cite{synge1966escape,novikov1973black,chandrasekhar1998mathematical,stuchlik2018light,grenzebach2015photon,atamurotov2013shadow,vagnozzi2019hunting,ma2021shadow}. In the last few years, the subject of black hole shadow aroused physicists' wide attentions. As we all know, the shadow of black hole is determined by light rays that fall into an event horizon, which depends on the black hole parameters and the interaction between light and other fields. Drummond and Hathrell \cite{drummond1980qed} first found that the superluminal photon propagation in gravitational backgrounds due to vacuum polarisation in QED induces interactions between the electromagnetic field and gravitational field, which has also been studied in Ref. \cite{daniels1994faster,daniels1996faster,cai1998propagation}. What's more, since light is actually a kind of electromagnetic wave, Ref. \cite{ritz2009weyl} has shown that coupling with Weyl tensor breaks the universal relation with the U(1) central charge observed at leading order. Then Songbai Chen $et$ $al$. studied the strong gravitational lensing for the photons coupled to Weyl tensor in black hole spacetimes \cite{chen2015strong,chen2017strong} and Yang Huang first discussed the double shadow of the black hole as
  photons couple to the Weyl tensor in Ref. \cite{huang2016double}. Recently, the coupling of electromagnetic field and gravitational field has been studied widely in Refs. \cite{zhang2020optical,chen2020polarization}. It indicates that the the interaction between electromagnetic field and gravitational field changes the path of photons propagation and leads to the birefringence phenomenon of light. 
  
  On the other hand, the discovery of gravity waves \cite{abbott2016observation} related to compact objects such as black holes, which is one of the outstanding discoveries, has provided us with a new window to understand the universe. It is well known that black hole perturbations result in the emission of gravitational waves, which are characterized by complex frequencies called quasinormal modes (QNMs). And QNMs have been investigated by using other analytic and numerical methods \cite{schutz1985black,ferrari1984new,chandrasekhar1975quasi,leaver1985analytic,horowitz2000quasinormal,iyer1987black,konoplya2003quasinormal,berti2009quasinormal,aragon2020perturbative}. What's more, a connection between the shadow of black hole and QNMs has attracted a lot of attention in recent years. Cardoso $et$ $al$. pointed out that in the eikonal limit, the real part of QNMs is connected to the last circular null geodesic in Ref. \cite{cardoso2009geodesic}, and Stefanov $et$ $al$. found a connection between the black hole QNMs and the strong gravitation lensing in the strong field regime \cite{stefanov2010connection}. Then in a recent work \cite{jusufi2020quasinormal}, Jusufi  explored the relationship between the shadow of nonrotating black hole and QNMs, and pointed out that in the eikonal limit the real part of QNMs is inversely proportional to the shadow radius, which is also applicable in the background of general rotating black holes \cite{jusufi2020connection,jusufi2020quasinormal1,liu2020shadow,ghasemi2020shadow,yang2021relating}. Therefore, it is very fair and motivating to connect these two properties as they open a new arena in black hole physics. And generalizing this correspondence to a more general case to test its domain of validation is still crucial.
  
  Our aim in this paper is to explore how the shadow of a 4D Einstein-Gauss-Bonnet black hole changes as photons couple to the Weyl tensor. And with above results in mind, we investigate the connection between the shadow radius and QNMs, which  is astrophysically relevant and has never been discussed. It is well known that in GB gravity, spherically symmetric black hole solutions exist only in higher dimensions due to the Gauss-Bonnet action does not contribute to the dynamics of the four-dimensional spacetime. Recently, a general covariant GB modified gravity in four dimensions was proposed by rescaling the GB coupling parameter $\hat{\alpha} \to\ \frac{\alpha}{D-4}$ which completed the missing piece of the EGB gravity \cite{glavan2020einstein}. Such theory can bypass the Lovelock’s theorem and avoid Ostrogradsky instability. A novel 4D static and spherically symmetric black hole solution was obtained while taking the limit $ D\to\ 4$. However, since the publication of the paper \cite{glavan2020einstein}, there have appeared several works \cite{ai2020note,shu2020vacua,arrechea2021inconsistencies,arrechea2020yet} debating that the procedure of taking $ D\to\ 4$ limit in \cite{glavan2020einstein} may not be consistent. On the other hand, some approaches have been proposed either by performing a regularized Kaluza–Klein reduction \cite{lu2020horndeski,kobayashi2020effective} or by introducing a counter term into the action \cite{fernandes2020derivation,hennigar2020taking} to circumvent the issues of the novel 4D EGB gravity. Moreover, a proposal has been suggested in \cite{aoki2020consistent}, which focuses on temporal diffeomorphism breaking instead of the inclusion of a scalar degree of freedom. Nevertheless, the black-hole solution of \cite{glavan2020einstein} also satisfies the field equations of the well-defined theory suggested in \cite{aoki2020consistent}. Therefore, the static and spherically symmetric black hole solution itself is meaningful and worthy of studying more deeply. Then, the charged \cite{fernandes2020charged} and rotating \cite{kumar2020rotating} analogues were obtained for new 4D EGB theory. What's more, very recently a lot of work has been made, which involved thermodynamics of black hole \cite{singh2020thermodynamics}, charged particles and epicyclic motions around 4D EGB black hole \cite{shaymatov2020charged} and the motions of spinning particles \cite{zhang2020spinning}.

  This paper is organized as follows. In Sec.~\ref{2}, we derive equations of motion for the photons coupled to the Weyl tensor in a 4D Einstein-Gauss-Bonnet black hole spacetime. Then in Sec.~\ref{3}, we further explore the change of the black hole shadow as photons couple to the Weyl tensor. Sec.~\ref{4} is devoted to studying the QNMs of scalar fields in a 4D Einstein-Gauss-Bonnet black hole spacetime by using the sixth-order WKB method. After that, the connection between the shadow radius and QNMs is investigated in Sec.~\ref{5}. Finally, we comment on our results in Sec.~\ref{6}

\section{EQUATION OF MOTION FOR THE PHOTONS COUPLED TO WEYL TENSOR}\label{2}
   In this section, we begin with the action of the electromagnetic field coupled to Weyl tensor in a 4D Einstein-Gauss-Bonnet black hole spacetime, which can be expressed as \cite{glavan2020einstein,ritz2009weyl}
  \begin{equation}\label{eq:action}
  	 S=\frac{1}{16\pi G}\int d^D x\sqrt{-g}[R+\frac{\alpha}{D-4} \mathcal{L}_{GB}]+\int d^D x\sqrt{-g}\mathcal{L}_{m},
  \end{equation}
  where the four-dimensional case is defined as the $D\to\ 4$ limit and $\mathcal{L}_{GB}=R_{\mu\nu}^{\rho\sigma} R_{\rho\sigma}^{\mu\nu}-4 R_{\mu}^{\nu}R_{\nu}^{\mu}+R^2 $ is the Gauss-Bonnet invariant. $\mathcal{L}_{m}=-\frac{1}{4}(F_{\mu\nu}F^{\mu\nu}-4\lambda C^{\mu\nu\rho\sigma}F_{\mu\nu}F_{\rho\sigma})$ is the Lagrangian density of the electromagnetic field coupled to the Weyl tensor in the D-dimensional spacetime and  $F_{\mu\nu}=\partial_{\mu}A_{\nu}-\partial_{\nu}A_{\mu}$ is the usual electromagnetic tensor. $C_{\mu\nu\rho\sigma}$ is the Weyl tensor with a form $C_{\mu\nu\rho\sigma}=R_{\mu\nu\rho\sigma}-\frac{2}{(D-2)}(g_{\mu[\rho}R_{\sigma]\nu}-g_{\nu[\rho}R_{\sigma]\mu})+\frac{2}{(D-1)(D-2)}Rg_{\mu[\rho}g_{\sigma]\nu}$. And the photons coupled to the Weyl tensor can be characterized by the coupling parameter $\lambda$. Now considering a four-dimensional spacetime, Varying the action (\ref{eq:action}) with respect to $A_{\mu}$, we can obtain the corrected  Maxwell equation  
  \begin{equation}\label{eq:Maxwell}
  	\nabla_{\mu}(F^{\mu\nu}-4\lambda C^{\mu\nu\rho\sigma}F_{\rho\sigma})=0.
  \end{equation}

  From the above corrected Maxwell equation (\ref{eq:Maxwell}), one can obtain the equation of motion of the coupled photons by the geometric optics approximation. Under this approximation, $\lambda_{c}<\lambda<L$, where $\lambda$ is the wavelength of photon, $L$ is a typical curvature scale, and $\lambda_{c}$ is the electron Compton wavelength \cite{daniels1994faster,daniels1996faster,cai1998propagation,shore2002faster,cho1997faster,de2000light,dalvit2001one}. In this method, the electromagnetic field strength can be written as
  \begin{equation}\label{eq:strength}
  	F_{\mu\nu}=f_{\mu\nu}\mathit{e}^{\mathit{i}\theta},
  \end{equation}
  where $f_{\mu\nu}$ is a slowly varying amplitude and $\theta$ is a rapidly varying phase. The wave vector is defined as $k_{\mu}=\partial_{\mu}\theta$, which corresponds to the photon momentum in the quantum partical interpretation. According to the Bianchi identity, the amplitude $f_{\mu\nu}$ has the form $f_{\mu\nu}=k_{\mu}a_{\nu}-k_{\nu}a_{\mu}$, where $a_{\mu}$ can be interpreted as the polarization vector of photon and satisfies $k_{\mu}a^{\mu}=0$. Inserting Eq. (\ref{eq:strength}) into Eq. (\ref{eq:Maxwell}) and using the relationship above, we arrive at
  \begin{equation}\label{eq:motion}
  	k_{\mu}k^{\mu}a^{\nu}+8\lambda C^{\mu\nu\rho\sigma}k_{\sigma}k_{\mu}a_{\rho}=0.
  \end{equation}
  Obviously, the coupling between Wely tensor and electromagnetic field affects the propagation of the coupled photons in the background spacetime.
  
  Let us now consider the 4D Einstein-Gauss-Bonnet black hole obtained by Glavan and Lin in \cite{glavan2020einstein}, with the metric form 
  \begin{equation}\label{eq:metric}
  	 \mathrm{d}s^{2}=-f\left(r\right)\mathrm{d}t^{2}+f\left(r\right)^{-1}\mathrm{d}r^{2}+r^{2}\mathrm({d}\theta^{2}+sin^{2}{\theta}\mathrm{d}\phi^{2})
  \end{equation} 
  where
  \begin{equation}
  	f(r)=1+\frac{r^{2}}{2\alpha}(1\pm\sqrt{1+\frac{8\alpha M}{r^{3}}}).
  \end{equation}
  Here, the sign $\pm$ refers to the two different branches, and we consider only the negative branch because it gives a Schwarzschild-like solution in the asymptotically limit \cite{glavan2020einstein,clifton2020observational,guo2020innermost,konoplya2020quasinormal,zhang2020greybody,yang2020weak}.
  
  Now introduce a local orthonormal frame. The appropriate basis 1-forms are $e^{a}(a=0,1,2,3)$ with
  \begin{equation}
  	e^{0}=\sqrt{f}dt,\quad\quad e^{1}=\frac{1}{\sqrt{f}}dr,\quad\quad e^{2}=rd\theta,\quad\quad e^{3}=rsin{\theta}d\phi.
  \end{equation}
  Introducing the notation $U{_{ab}^{01}}=\delta_a^0 \delta_b^1-\delta_a^1\delta_b^0$ \cite{drummond1980qed,daniels1994faster,daniels1996faster,cai1998propagation}, etc., the Weyl tensor can be simplified as
  \begin{equation}\label{eq:weyl}
  	C_{abcd}=\mathcal{A}(2U{_{ab}^{01}}U{_{cd}^{01}}-U{_{ab}^{02}}U{_{cd}^{02}}-U{_{ab}^{03}}U{_{cd}^{03}}+U{_{ab}^{12}}U{_{cd}^{12}}+U{_{ab}^{13}}U{_{cd}^{13}}-2U{_{ab}^{23}}U{_{cd}^{23}})
  \end{equation}
  with
  \begin{equation}
	\mathcal{A}=-\frac{M(r^3 +2M\alpha)\sqrt{1+\frac{8M\alpha}{r^3}}}{(r^3+8M\alpha)^2}.
  \end{equation}
  To solve the equation of motion of photon, it is convenient to introduce the following linear combination of momentum components \cite{drummond1980qed,daniels1994faster,daniels1996faster,cai1998propagation}
  \begin{equation}
  		l_b=k^a U{_{ab}^{01}},\quad\quad m_b=k^a U{_{ab}^{02}},\quad\quad r_b=k^a U{_{ab}^{03}},
  \end{equation}
  together with the dependent combinations
 \begin{equation}
 	p_b=k^a U{_{ab}^{12}},\quad\quad q_b=k^a U{_{ab}^{13}},\quad\quad n_b=k^a U{_{ab}^{23}}.
  \end{equation}
  The equation of motion for the coupled photons (\ref{eq:motion}) can be rewritten as a set of equations for the independent polarisation components $a\cdot l$, $a\cdot m$ and $a\cdot n$. Plugging Eq. (\ref{eq:weyl}) into Eq. (\ref{eq:motion}), we therefore arrive at
  \begin{equation}\label{eq:motion matrix}
  	\begin{pmatrix}
  		K_{11} & 0 & 0 \\ K_{21} & K_{22} & K_{23} \\ 0 & 0 & K_{33}
   	\end{pmatrix}
   \begin{pmatrix}
   	a\cdot l \\ a\cdot m \\ a\cdot n
   \end{pmatrix} = 0 
  \end{equation}
  with 
  \begin{align}
  	K_{11} & =(1-16\lambda\mathcal{A})(-k^0 k^0+k^1 k^1)+(1+8\lambda\mathcal{A})(k^2 k^2+k^3 k^3),\notag \\
  	K_{21} & =-24\lambda\mathcal{A} k^1 k^2,\notag \\
 	K_{22} & =(1+8\lambda\mathcal{A})(-k^0 k^0+k^1 k^1+k^2 k^2+k^3 k^3), \\
 	K_{23} & =24\lambda\mathcal{A} k^0 k^3,\notag \\
 	K_{33} & =(1+8\lambda\mathcal{A})(-k^0 k^0+k^1 k^1)+(1-16\lambda\mathcal{A})(k^2 k^2+k^3 k^3).\notag 
  \end{align}
  The condition of Eq. (\ref{eq:motion matrix})  with the non-zero solution is $K_{11}K_{22}K_{33}=0$. The first root $K_{11}=0$ leads to the modified light cone
  \begin{equation}\label{eq:ppl}
  	(1-16\lambda\mathcal{A})(-k^0 k^0+k^1 k^1)+(1+8\lambda\mathcal{A})(k^2 k^2+k^3 k^3)=0,
  \end{equation}
  which corresponds to the case of the polarisation vector $a_{\mu}=\lambda l_{\mu}$. The second root $K_{22}=0$ corresponds to an unphysical polarisation and should be neglected. The third root is $K_{33}=0$, i.e.,
  \begin{equation}\label{eq:ppr}
  	(1+8\lambda\mathcal{A})(-k^0 k^0+k^1 k^1)+(1-16\lambda\mathcal{A})(k^2 k^2+k^3 k^3)=0,
  \end{equation}
  which means that the vector $a_{\mu}=\lambda n_{\mu}$.
  
  The above discussion shows that the light cone condition depends not only on the coupling between the photon and the Weyl tensor, but also on the polarization of light. We know from Eqs. (\ref{eq:ppl}) and (\ref{eq:ppr}) that the velocities of the photons for the two polarizations are different, i.e., the phenomenon of gravitational birefringence. Moreover, the light cone conditions (\ref{eq:ppl}) and (\ref{eq:ppr}) imply that instead of following geodesic in the original metric, the coupled photons follow null geodesics of the effective metric $\gamma_{\mu\nu}$, i.e., $\gamma^{\mu\nu}k_{\mu}k_{\nu}=0$ \cite{preuss2004astronomical}. The effective metric for the coupled photon can be expressed as
   \begin{equation}\label{eq:effective metric}
  	\mathrm{d}s^{2}=-f\left(r\right)\mathrm{d}t^{2}+f\left(r\right)^{-1}\mathrm{d}r^{2}+r^{2}W(r)^{-1}\mathrm({d}\theta^{2}+sin^{2}{\theta}\mathrm{d}\phi^{2}).
  \end{equation}
  The quantity $W(r)$ is
  
  \begin{equation}
 	W(r)=\frac{r^6 \sqrt{1+\frac{8M\alpha}{r^3}}+Mr^3(8\alpha\sqrt{1+\frac{8M\alpha}{r^3}}-8\lambda)-16M^2\alpha\lambda}{r^6\sqrt{1+\frac{8M\alpha}{r^3}}+32M^2\alpha\lambda+Mr^3(8\alpha\sqrt{1+\frac{8M\alpha}{r^3}}+16\lambda)},
  \end{equation}
  for photon with the polarization along $l_{\mu} (PPL)$, it is 
  
  \begin{equation}
 	W(r)=\frac{r^6\sqrt{1+\frac{8M\alpha}{r^3}}+32M^2\alpha\lambda+Mr^3(8\alpha\sqrt{1+\frac{8M\alpha}{r^3}}+16\lambda)}{r^6\sqrt{1+\frac{8M\alpha}{r^3}}+Mr^3(8\alpha\sqrt{1+\frac{8M\alpha}{r^3}}-8\lambda)-16M^2\alpha\lambda},
  \end{equation}
  for photon with the polarization along $n_{\mu} (PPN)$ \cite{huang2016double}.
\section{WEYL CORRECTIONS TO SHADOW RADIUS}\label{3}
  In this subsection, we will discuss the shadow radius of black hole as photons couple to the Weyl tensor. In the effective metric (\ref{eq:effective metric}), there exist two conserved quantites energy $E$ and angular momentum $L$ as follow 
  \begin{equation}
  	E=f(r)\dot{t},\quad\quad L=r^{2}sin^{2}{\theta}W(r)^{-1}\dot{\phi},
  \end{equation} 
  where the dot over a symbol is the differentiation with respect to an affine parameter $\beta$. Using the condition $\gamma^{\mu\nu}k_{\mu}k_{\nu}=0$ and $k_{\mu}=\frac{dx_{\mu}}{d\beta}$, one can obtain the equations of motion
  \begin{gather}
  	\frac{r^4\dot{r}^2}{W(r)^2}=R(r), \\
  	\frac{r^4\dot{\theta}^2}{W(r)^2}=\Theta(\theta).
  \end{gather}
  Here, $R(r)$ and $\Theta(\theta)$ are given by
  \begin{gather}
  	R(r)=\frac{E^2 r^4}{W(r)^2}-\frac{(\mathcal{Q}+L^2)r^2f(r)}{W(r)}, \\
  	\Theta(\theta)=\mathcal{Q}-L^2cot^{2}{\theta},
  \end{gather}
  with $\mathcal{Q}$ denoting the Carter constant \cite{carter1968global}. To determine the geometric sharp of the shadow of the effective metric, using the unstable condition
  \begin{equation}\label{eq:unstable condition}
  	R(r)=0,\quad\quad \frac{dR(r)}{dr}=0,\quad\quad \frac{d^2R(r)}{dr^2}>0,
  \end{equation}
   one can obtain that
  \begin{equation}\label{eq:photon sphere orbit}
  	2f(r)W(r)-rf'(r)W(r)-rf(r)W'(r)=0.
  \end{equation}
  By solving this equation, one can determine the radius of the photon sphere $r_{ps}$. In particular, combining Eq. (\ref{eq:unstable condition}) with celestial coordinates \cite{vazquez2003strong}
  \begin{equation}
    x=\lim_{r_i\to\infty}(-r^2_isin{\theta_i}\frac{d\phi}{dr})\vert_{(r_i,\theta_i)},\quad\quad y=\lim_{r_i\to\infty}(r^2_i\frac{d\theta}{dr})\vert_{(r_i,\theta_i)},
  \end{equation}
   one can express the radius of the shadow $R_{sh}$ of the effective metric by the simple relation
  \begin{equation}
  	R_{sh}=\sqrt{x^2+y^2}=\frac{r_{ps}}{\sqrt{W(r_{ps})f(r_{ps})}},
  \end{equation}
while taking the limit $r_i\to\infty$ according to the location of the  observer and assuming the observer is situated in the equatorial plane ($\theta_i=\frac{\pi}{2}$).

\begin{figure}
	\centering
	\includegraphics[width=8.2cm]{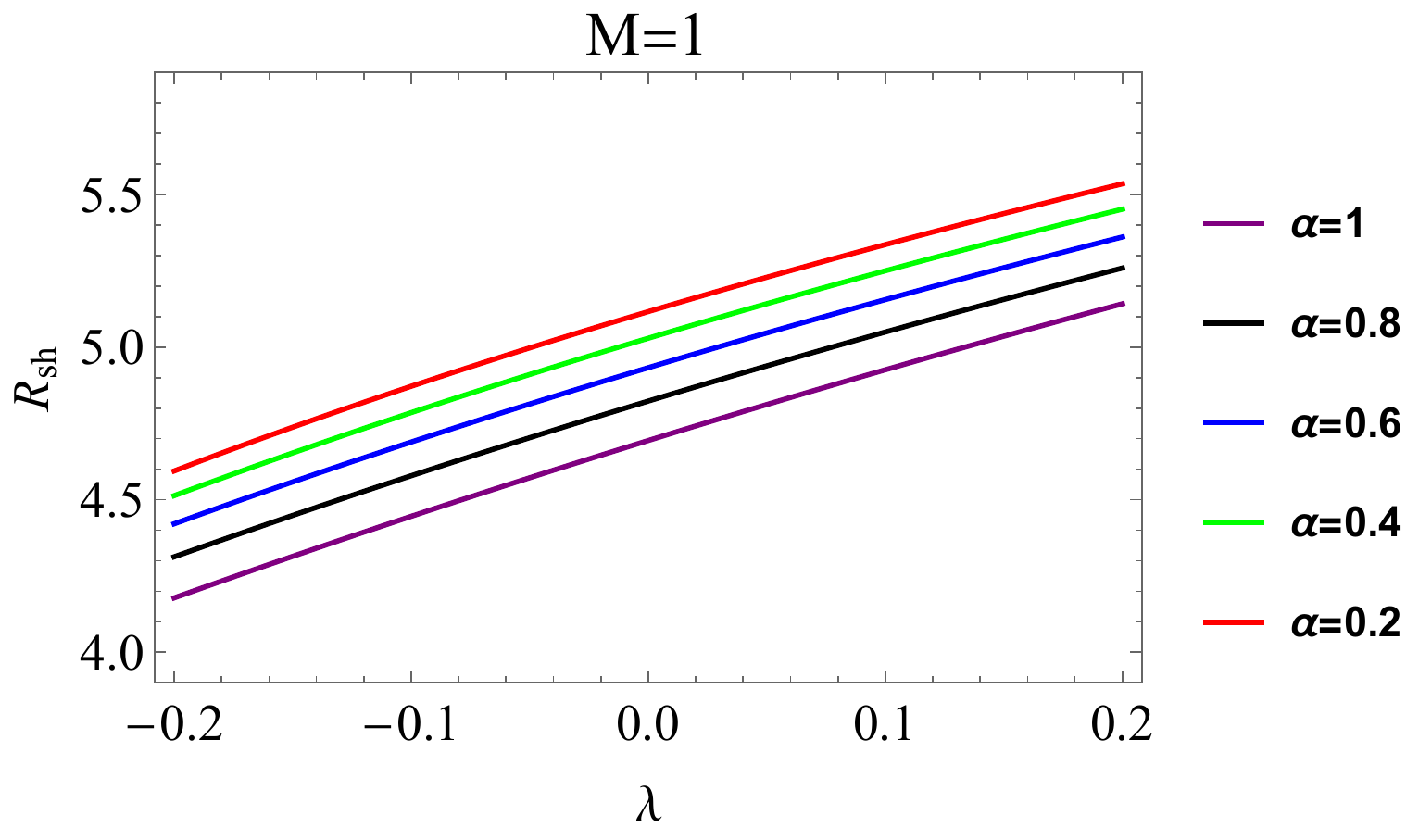}
	\includegraphics[width=6.6cm]{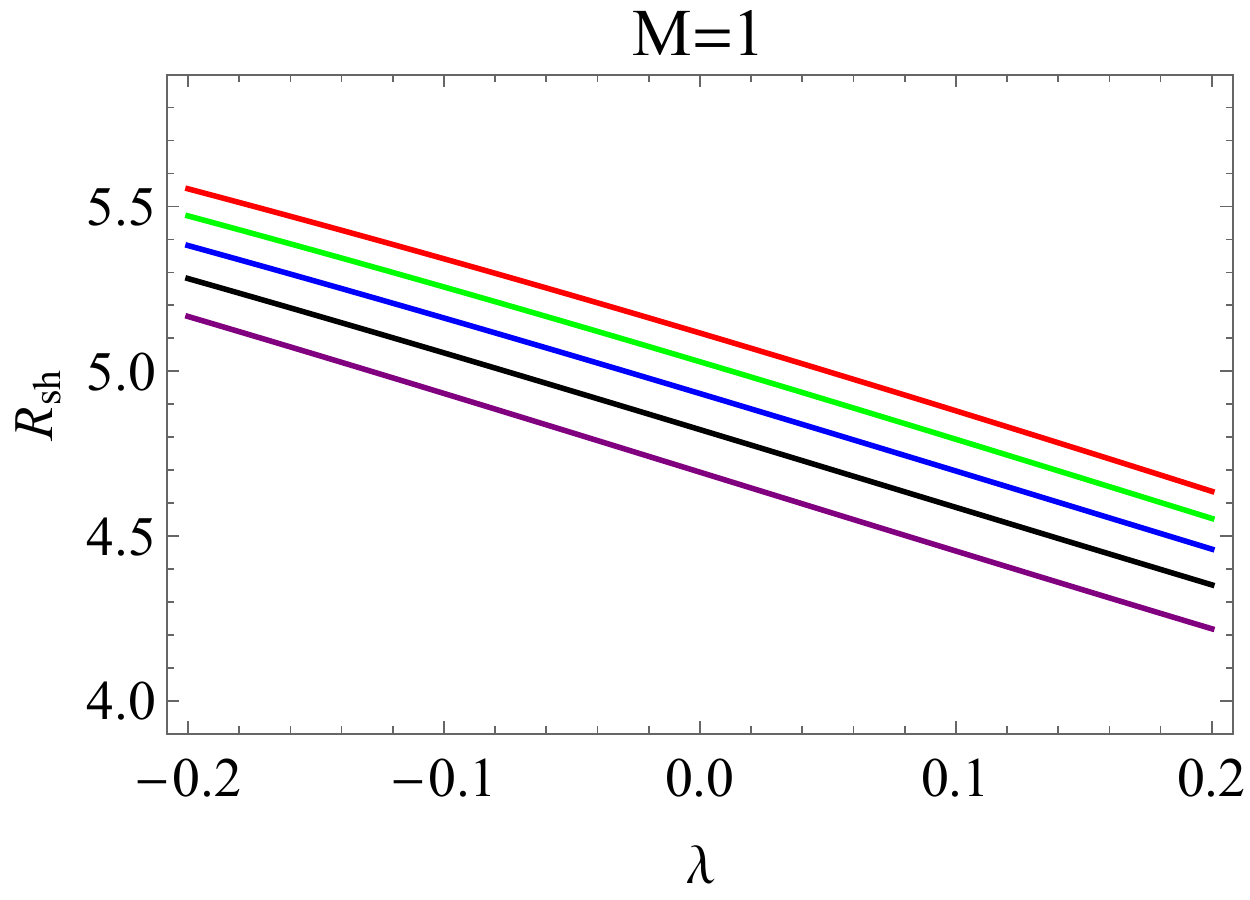}
	\caption{Variation of the shadow radius $R_{sh}$ with the coupling parameter  $\lambda$ in a 4D Einstein-Gauss-Bonnet black hole spacetime. The left and the right are for PPL and PPN, respectively.}
	\label{shadow1}	
\end{figure}

\begin{figure}
  	\centering
  	\includegraphics[width=8.2cm]{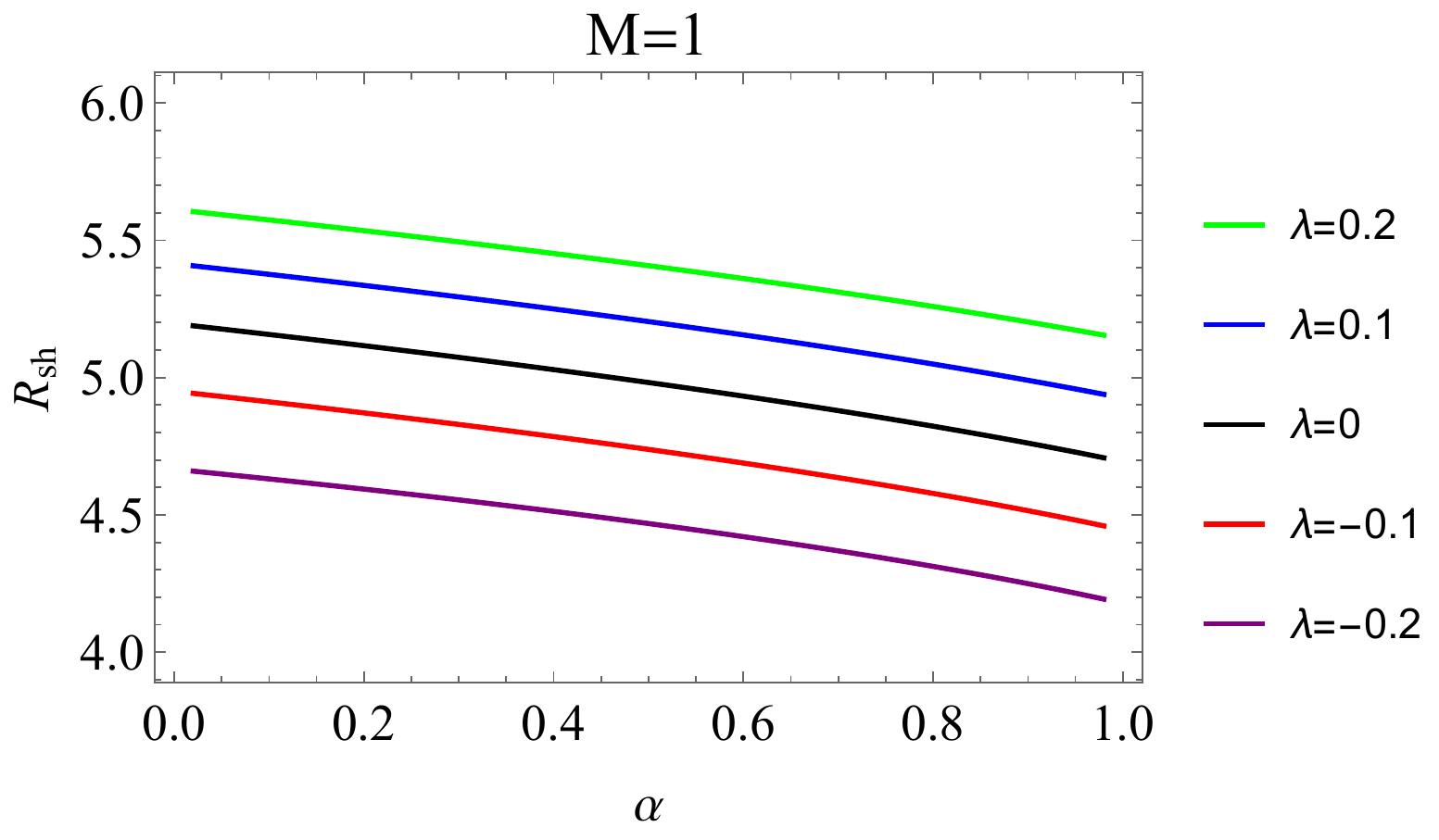}
  	\includegraphics[width=6.6cm]{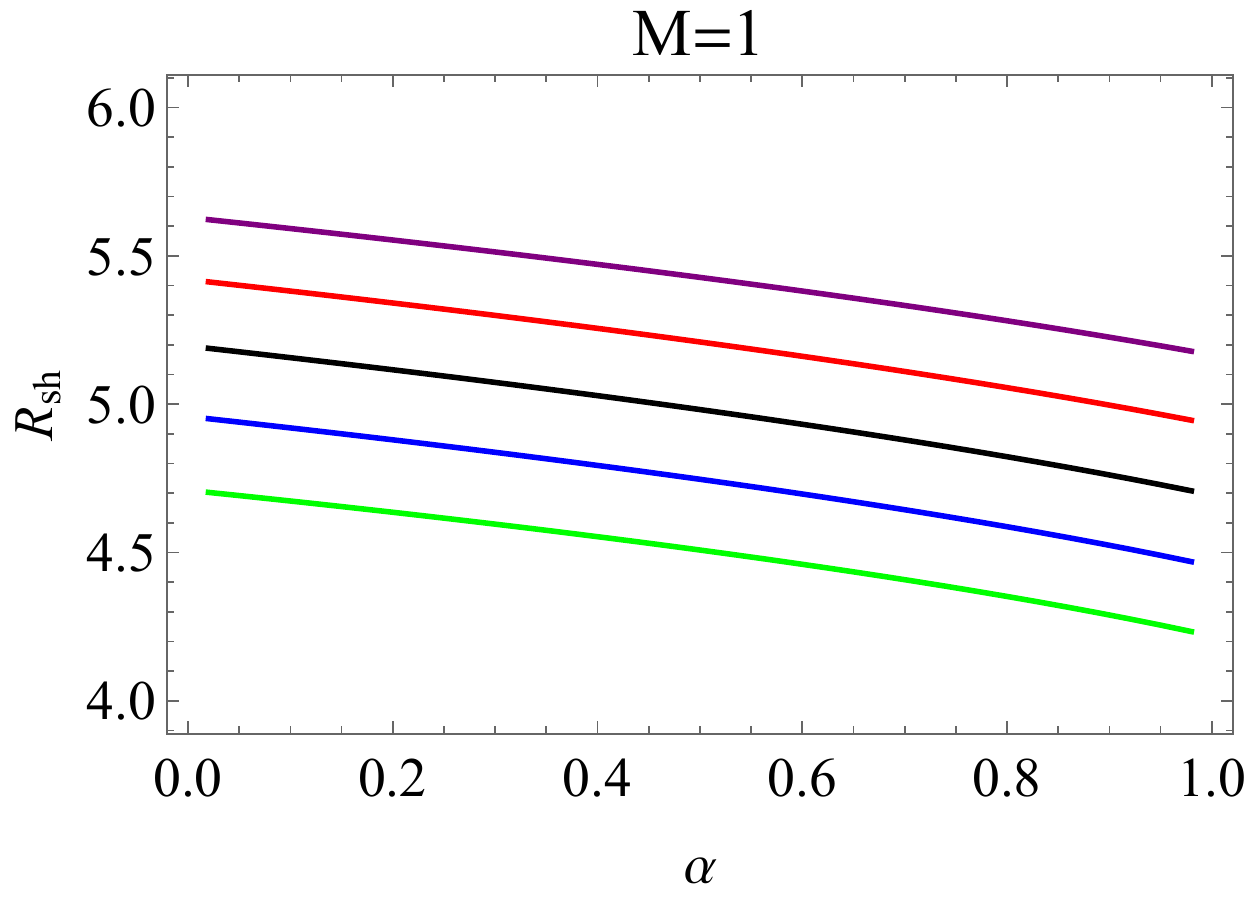}
  	\caption{Variation of the shadow radius $R_{sh}$ with the GB coupling parameter  $\alpha$ in a 4D Einstein-Gauss-Bonnet black hole spacetime. The left and right are for PPL and PPN, respectively.}
  	\label{shadow2}	
  \end{figure}

\begin{table*}
	\centering
	\begin{tabular}{cccccccccccccccccccccc}
		\hline
		\hline
		${}$ & \multicolumn{6}{c}{$r_{+}$} && \multicolumn{6}{c}{$r_{ps}/R_{sh}$(PPL)} &&&& \multicolumn{5}{c}{$r_{ps}/R_{sh}$(PPN)} \\
		\hline
		$\alpha$ &&&$ $&&&&& $\lambda=0.2$ && $\lambda=0$ &&$\lambda=-0.2$  &&&&& $\lambda=0.2$ && $\lambda=0$ && $\lambda=-0.2$  \\
		\hline
		0.2 &&& 1.894 &&&&& 3.155/5.535 && 2.907/5.116 && 2.639/4.594 &&&&& 2.675/4.636 && 2.907/5.116 && 3.179/5.554 \\
		0.4 &&& 1.775 &&&&& 3.049/5.452 && 2.803/5.029 && 2.550/4.513 &&&&& 2.583/4.553 && 2.803/5.029 && 3.075/5.471 \\
		0.5 &&& 1.707 &&&&& 2.992/5.407 && 2.747/4.982 && 2.500/4.469 &&&&& 2.532/4.508 && 2.747/4.982 && 3.017/5.428 \\
	    0.6 &&& 1.632 &&&&& 2.930/5.361 && 2.686/4.932 && 2.445/4.421 &&&&& 2.476/4.460 && 2.686/4.932 && 2.957/5.382 \\
		0.8 &&& 1.447 &&&&& 2.791/5.259 && 2.547/4.823 && 2.315/4.313 &&&&& 2.345/4.352 && 2.547/4.823 && 2.819/5.281 \\
		  1 &&& 1     &&&&& 2.622/5.142 && 2.372/4.694 && 2.140/4.178 &&&&& 2.170/4.219 && 2.372/4.694 && 2.654/5.166 \\
		\hline
		\hline
	\end{tabular}
	\caption{The event horizon $r_{+}$, photon sphere radius $r_{ps}$ and shadow radius $R_{sh}$ with different GB coupling constant $\alpha$ in a 4D Einstein-Gauss-Bonnet black hole spacetime.}
	\label{t0}
\end{table*}
   Considering that a photon should propagate continuously in the region outside the event horizon $r_{+}$, the coupling constant $\lambda$ must satisfy $r_{+}^6\sqrt{1+\frac{8M\alpha}{r_{+}^3}}+32M^2\alpha\lambda+Mr_{+}^3(8\alpha\sqrt{1+\frac{8M\alpha}{r_{+}^3}}+16\lambda)>0$ for PPL and  satisfy $r_{+}^6\sqrt{1+\frac{8M\alpha}{r_{+}^3}}+Mr_{+}^3(8\alpha\sqrt{1+\frac{8M\alpha}{r_{+}^3}}-8\lambda)-16M^2\alpha\lambda>0$ for PPN. With this constraint, setting $M=1$, the range of the GB coupling parameter  $\alpha$ is chosen in $0< \alpha\leq 1$ and the coupling constant $\lambda$ is limited in the range $\lambda \in[-0.2,0.2] $, such that the condition $r_{+}<r_{ps}<R_{sh}$ holds \cite{lu2019size}, seeing Table \ref{t0}. Obviously, due to the complex dependence of the equation (\ref{eq:photon sphere orbit}) on the GB coupling parameter $\alpha$ and the coupling constant $\lambda$, we can not get an analytical form for the photon sphere radius for the coupled photons, so the computational way is chosen. With the help of the numerical method, we can get the variation of $R_{sh}$ with parameter $\lambda$ of the photons coupled to the Weyl tensor and the GB coupling constant $\alpha$ for PPL and PPN, and the results are shown in Figs. \ref{shadow1} and \ref{shadow2}. 
  
\par
  In Fig. \ref{shadow1}, it tells us that with the increase of the coupling parameter $\lambda$, the shadow radius $R_{sh}$ for different values of $\alpha$ increases for PPL and decreases for PPN. However, as $\lambda \to\ 0 $, $W(r) \to\ 1$, the equation of circular photon orbits for PPL is the same as that for PPN, and the solution (\ref{eq:effective metric}) reduces to the 4D Einstein-Gauss-Bonnet black hole with the shadow investigated in \cite{panah2020charged}. The variation of the shadow radius with the GB coupling parameter $\alpha$ in our article is consistent with the result in \cite{panah2020charged}. In addition, one can find that the shadow depends not only on the properties of background black hole spacetime, but also on the polarization of the coupled photons. Moreover, with the increase of the GB coupling parameter $\alpha$,  the shadow radius $R_{sh}$ decreases for different polarizations of photons in Fig. \ref{shadow2}. The shadow radius $R_{sh}$ of PPL is different from that of PPN in Fig. \ref{shadow2} for fixed value of $\lambda$, which means the birefringence phenomenon occurs exactly in this case with non-vanishing $\lambda$. When $\alpha \to\ 0 $, the $R_{sh}$ reduces to the radius of the shadow for the photons coupled to Weyl tensor in a Schwarzschild black hole spacetime which has been studied in \cite{chen2015strong}. It indicates that with increase of $\lambda$, the shadow radius $R_{sh}$ increases for PPL and decreases for PPN in the Schwarzschild black hole spacetime, which is consistent with our result. We also find that the shadow radius in the 4D Einstein-Gauss-Bonnet black hole spacetime is always smaller than that in Schwarzschild black hole spacetime for the GB coupling constant $0 < \alpha\leq 1$. 
\section{QNMs OF SCALAR FIELD}\label{4}
   Considering a massless scalar field perturbation in the metric (\ref{eq:metric}), it satisfies the Klein-Gorden equation
  \begin{equation}\label{eq:Klein-Gorden}
  	\frac{1}{\sqrt{-g}}\partial_{\mu}(\sqrt{-g}g^{\mu\nu}\partial_{\nu}\Phi)=0.
  \end{equation}
  Involving a separation of variables, the function $\Phi$ for the scalar field is given in terms of the spherical harmonics 
  \begin{equation}\label{eq:spherical harmonics}
  	\Phi(t,r,\theta,\phi)=\frac{1}{r}e^{-i\omega t}Y_{l}(\theta,\phi)\Psi(r),
  \end{equation}
  in which $e^{-i\omega t}$ represents the time evolution of the field. Inserting Eq. (\ref{eq:spherical harmonics}) into Eq. (\ref{eq:Klein-Gorden}) and introducing a "tortoies" coordinate $dr_*=\frac{dr}{f(r)}$, we can show that the field perturbation equation is given by the Schrödinger wave-like equation
  \begin{equation}\label{eq:Schrödinger}
  	\frac{d^2\Psi}{dr_*^2}+(\omega^2-V_S(r))\Psi=0.
  \end{equation}

  Under the positive real part, QNMs, by definition, satisfy the following boundary condition,
  \begin{equation}
  	\Psi(r_*)=C_{\pm} \exp({\pm}i\omega r_*),\quad\quad r\to\infty
  \end{equation}
  where $\omega$ can be further written in terms of the real and imaginary parts, i.e., $\omega =\omega_R-i \omega_I$. The real and imaginary parts of QNMs represent the oscillation frequency and the decay rate, respectively. The effective potential $V_S(r)$ in Eq. (\ref{eq:Schrödinger}) can be written as
  
  \begin{equation}
  	V_S(r)=\left[1+\frac{r^{2}}{2\alpha}(1-\sqrt{1+\frac{8\alpha M}{r^{3}}})\right] \times \left[\frac{l(l+1)}{r^2}-\frac{2\alpha+r^3\sqrt{1+\frac{8\alpha M}{r^{3}}}}{\alpha r^3\sqrt{1+\frac{8\alpha M}{r^{3}}}}\right],
  \end{equation}
  \begin{figure} 
  	\centering
  	\includegraphics[width=7.4cm]{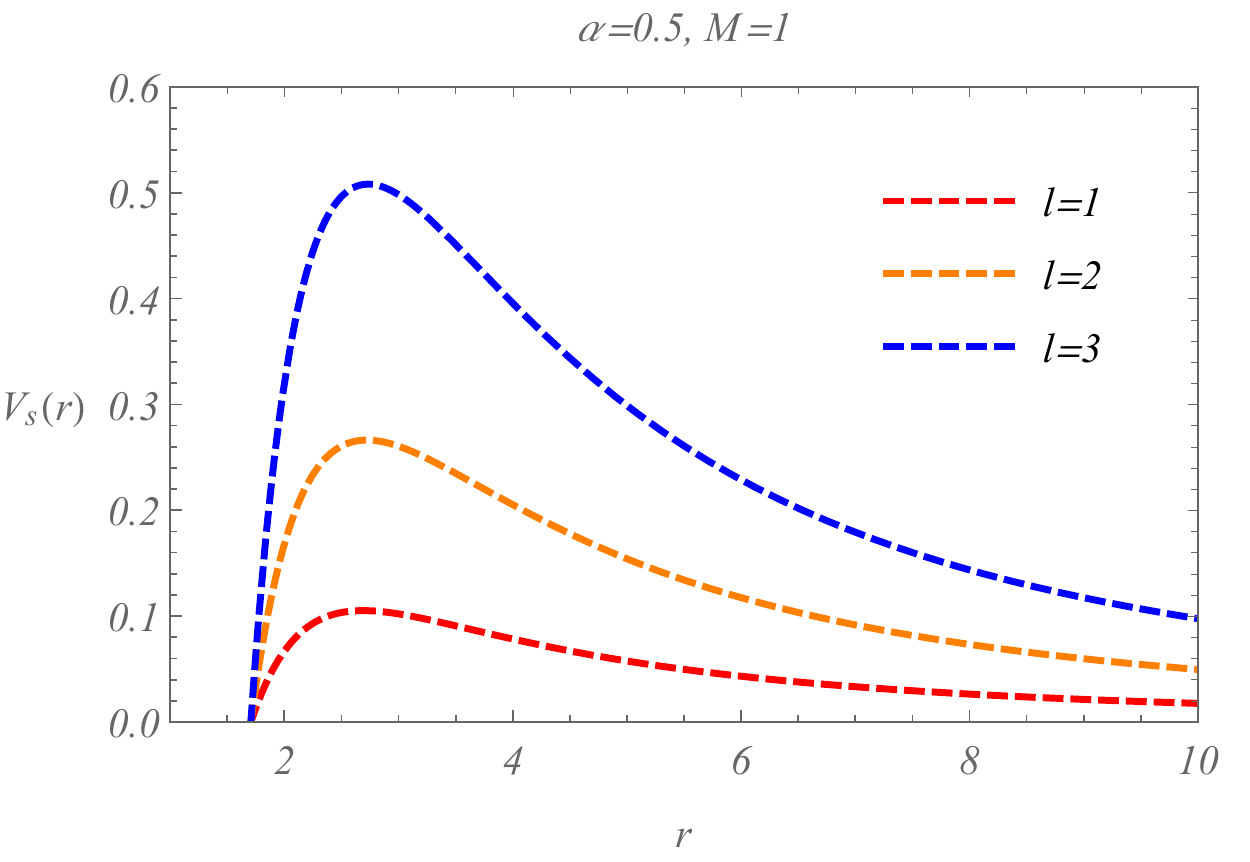}
  	\includegraphics[width=7.4cm]{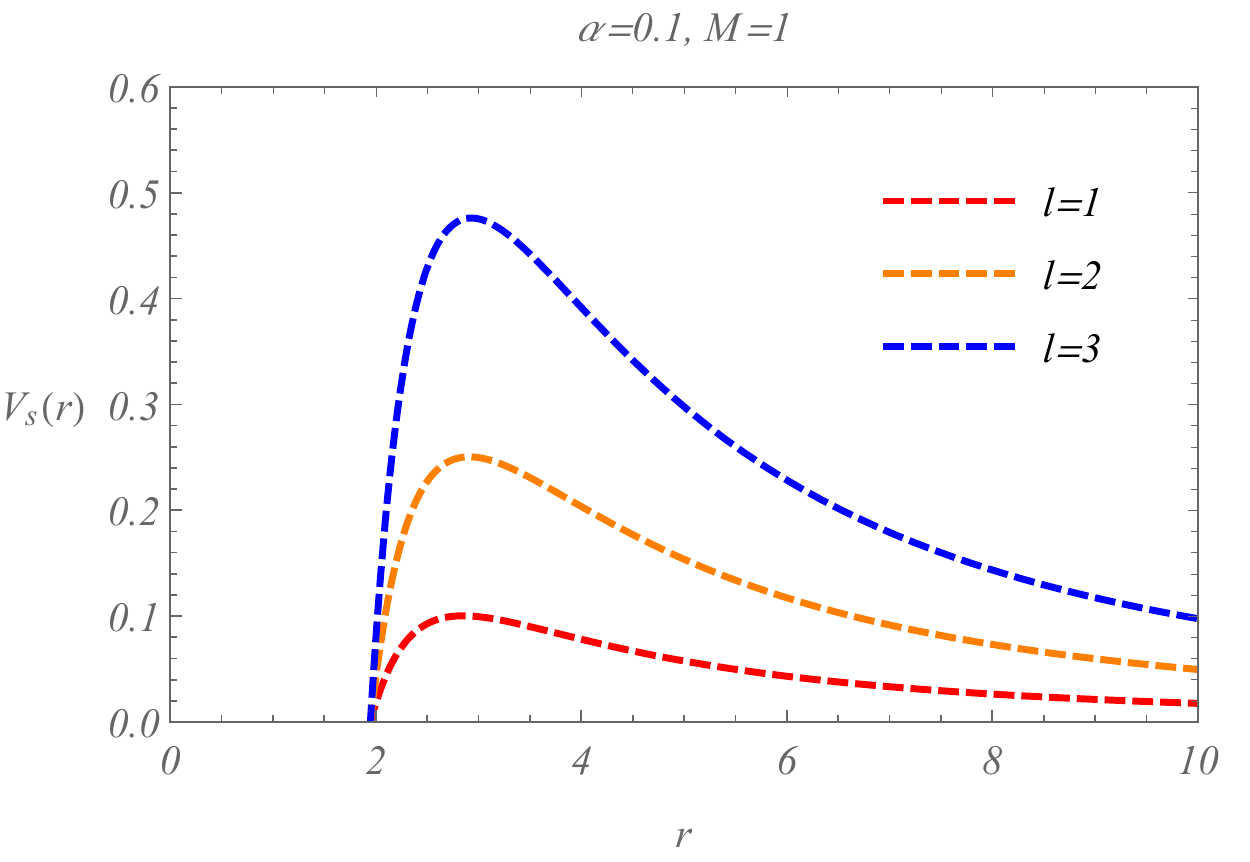}
  	\caption{The figures are the effective potentials of the scalar field perturbation $V_S$ for different values of the GB coupling constant $\alpha$. Changing the parameter $\alpha$ changes the height of the potential barrier.}
  	\label{effective potentials}	
  \end{figure}
  \begin{table}
  	\centering
  	\begin{tabular}{ccccccccccccccc}
  		\hline
  		\hline
  		spin 0 &&&& l=1, n=0 & &&&& l=2, n=0 & &&&& l=2, n=1 \\
  		\hline
  		$\alpha$ &&&& $\omega$(WKB) &&&&& $\omega$(WKB) &&&&& $\omega$(WKB) \\
  		\hline
  		$0.2$ &&&& $0.2977-0.0948 i$ &&&&& $0.4914-0.0936 i$ &&&&& $0.4736-0.2853 i$ \\
  		$0.4$ &&&& $0.3028-0.0912 i$ &&&&& $0.5000-0.0899 i$ &&&&& $0.4839-0.2734 i$ \\
  		$0.5$ &&&& $0.3057-0.0891 i$ &&&&& $0.5048-0.0877 i$ &&&&& $0.4892-0.2665 i$ \\
  		$0.6$ &&&& $0.3087-0.0867 i$ &&&&& $0.5099-0.0853 i$ &&&&& $0.4947-0.2588 i$ \\
  		$0.8$ &&&& $0.3154-0.0806 i$ &&&&& $0.5211-0.0795 i$ &&&&& $0.5057-0.2399 i$ \\
  		$  1$ &&&& $0.3227-0.0723 i$ &&&&& $0.5343-0.0711 i$ &&&&& $0.5146-0.2147 i$ \\
  		\hline
  		\hline
  	\end{tabular}
  	\caption{The real and imaginary parts of quasinormal frequencies of the scalar field in the 4D Einstein-Gauss-Bonnet black hole spacetime with different coupling constant $\alpha$.}
  	\label{t1}
  \end{table}
  where  $l$ denotes the multipole number. One can deduce from Fig. \ref{effective potentials} that the height of the potential barrier governed by the effective potential increases with multipole number $l$ and the GB coupling parameter $\alpha$. With the expression of the effective potential in hand, one can use the WKB approach to compute the QNM frequencies. This method was proposed by Schutz and Will \cite{schutz1985black}, then Iyer and Will extended the first order WKB formula to the third order \cite{iyer1987black}. In the present paper, we shall use the sixth-order WKB method which is described in \cite{konoplya2003quasinormal} for calculating QNMs. We have presented the values of the QNMs for the scalar perturbation in Table \ref{t1}.
  
  \begin{figure} 
  	\centering
  	\includegraphics[width=7.4cm]{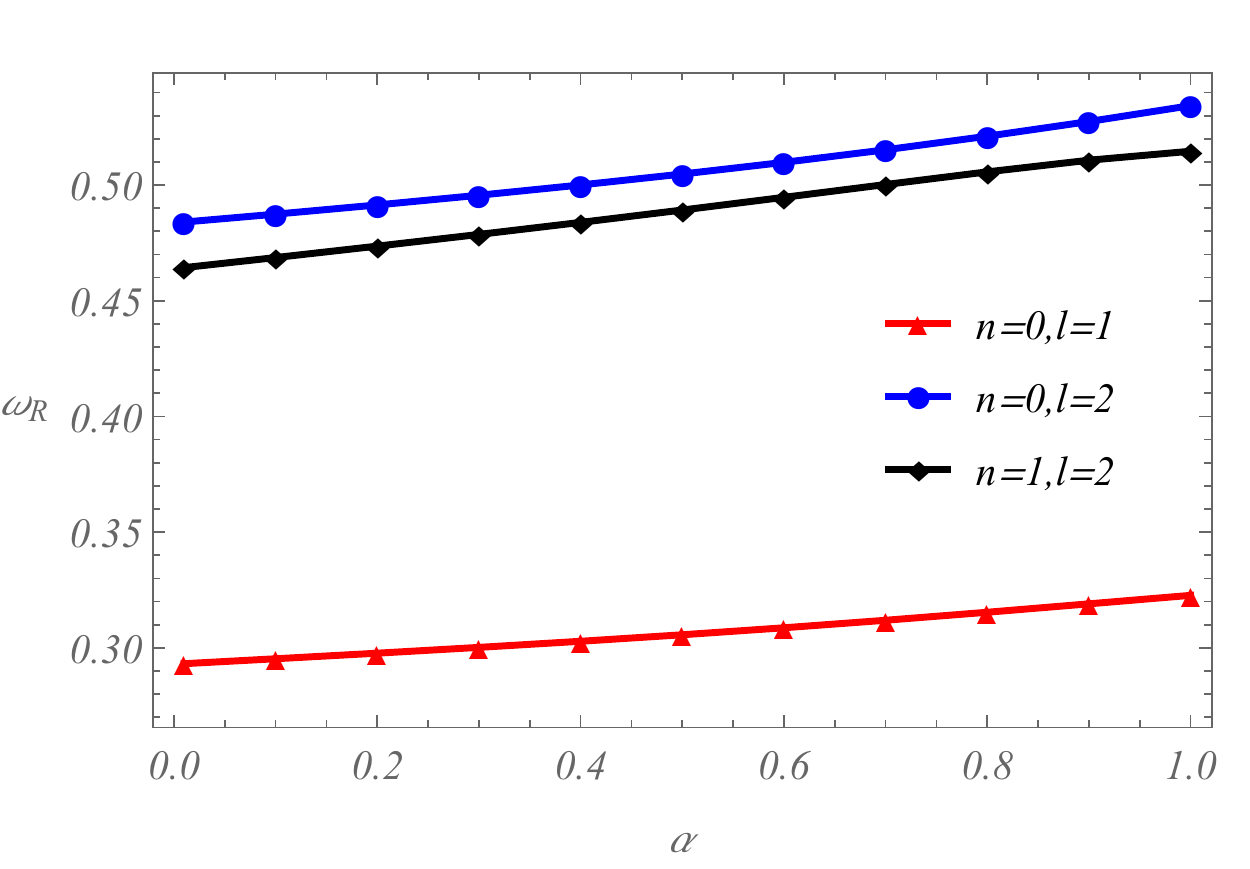}
  	\includegraphics[width=7.4cm]{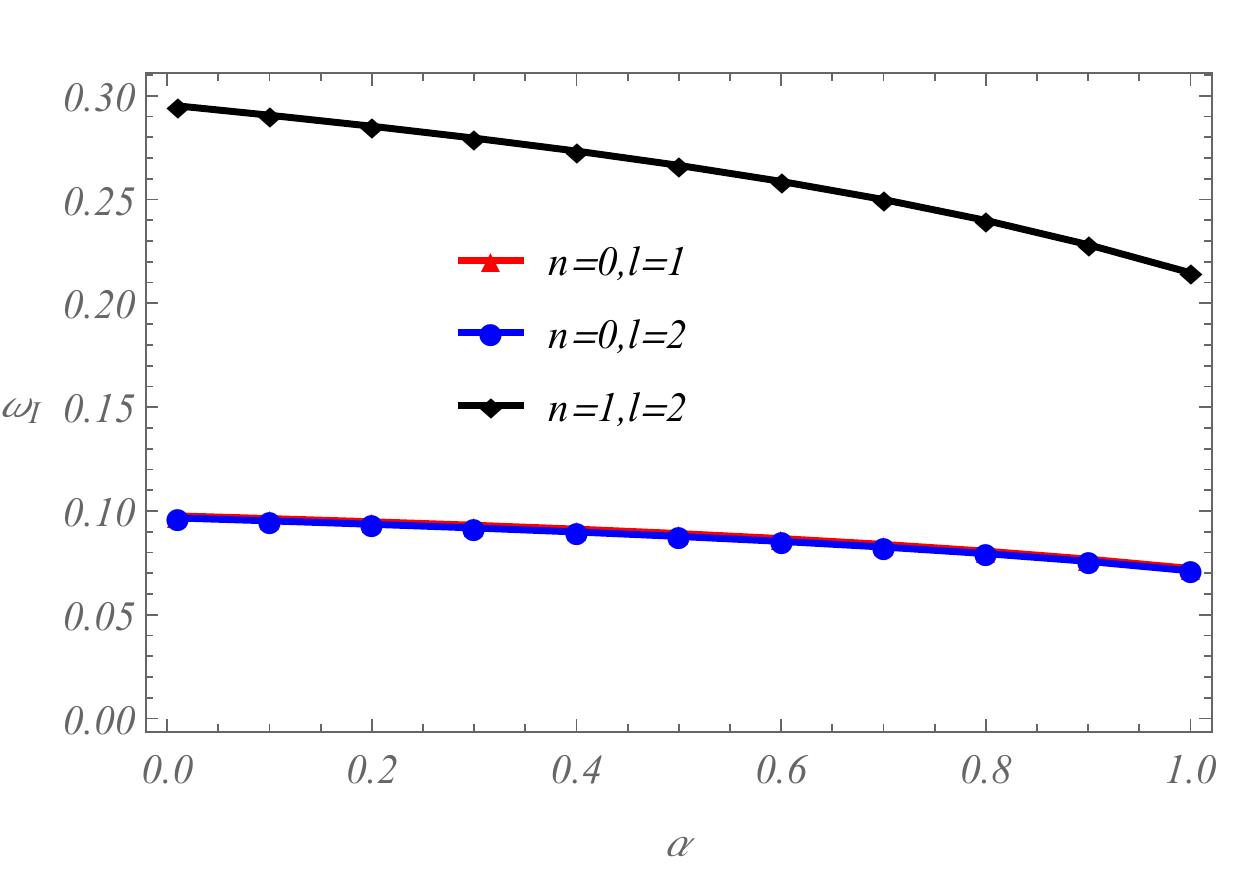}
  	\caption{Left panel: Plots of the real part of the QNMs versus the GB coupling parameter $\alpha$ for the scalar field. Right panel: Plots of the imaginary part of QNMs versus the GB coupling parameter $\alpha$. In both cases we have chosen $M=1$.}
  	\label{QNMs}
  \end{figure}
  From Table \ref{t1}, we find that with the increase of the GB coupling parameter $\alpha$, the real part of QNMs increases and the imaginary part decreases in absolute value. This particular effect can be clearly seen in Fig. \ref{QNMs}. Moreover, we also see that higher absolute values of $\omega_I$ are botained for the case $l=2$ and $n=1$, which means the scalar field perturbation damps more rapidly. It is worth noting that the WKB method works for $l>n$ (where $n$ is the overtone number), otherwise, the accuracy is worse. So that we have not calculated the QNMs for the fundamental mode $l=0,n=0$, but can use the Frobenius method \cite{konoplya2018massive}.

\section{CONNECTION BETWEEN THE SHADOW RADIUS AND QNMs}\label{5}
\begin{figure} 
  	\centering
  	\includegraphics[width=9.4cm]{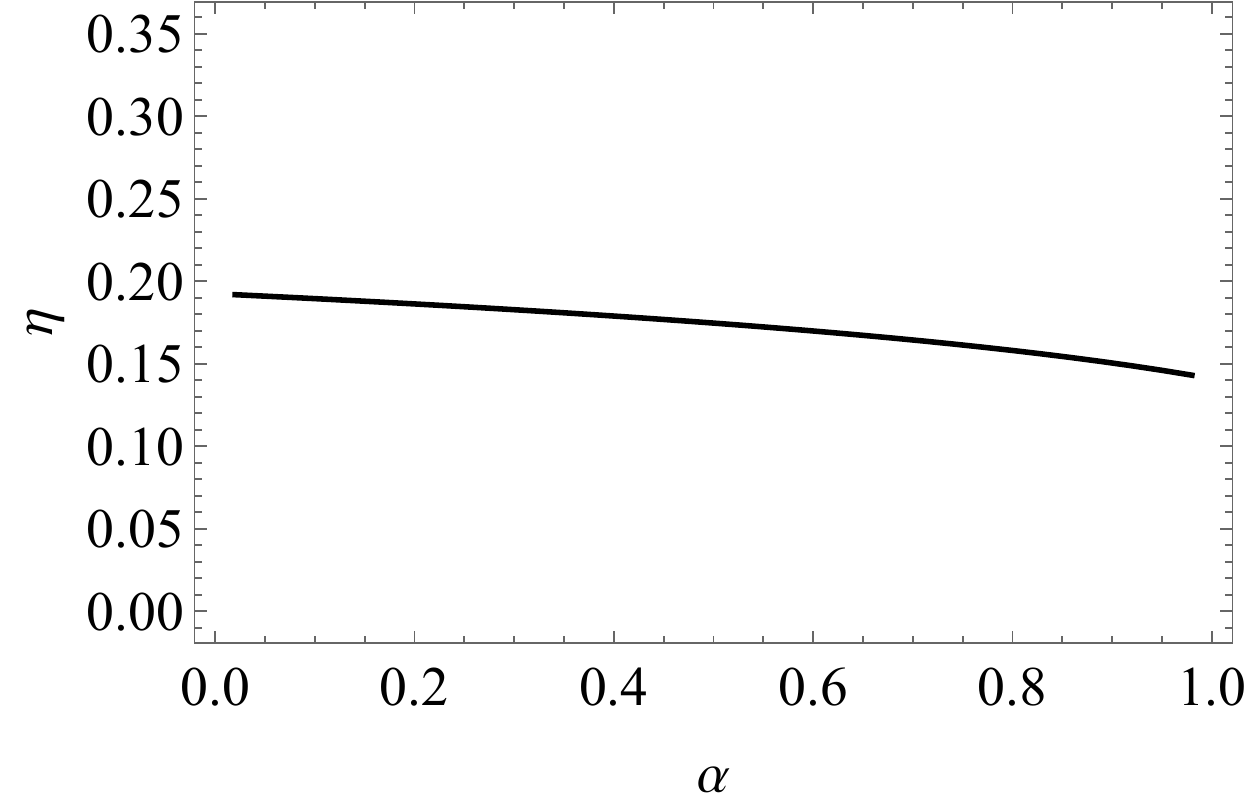}
  	\caption{ Plot of the  Lyapunov exponent $\eta$ versus the GB coupling parameter $\alpha$. In this cases we have chosen $M=1$.}
  	\label{the Lyapunov exponent}
  \end{figure}

   At high $l$, the WKB formula found in \cite{schutz1985black,iyer1987black,konoplya2003quasinormal} can be applied for finding quasinormal modes
   \begin{equation}
    \frac{Q_{0}(r_{0})}{\sqrt{2Q_{0}^{(2)}(r_{0})}}=i(n+1/2),
   \end{equation}
   where $Q_{0}^{(2)}=\frac{d^2Q_{0}}{dr_{*}^{2}}$ is evaluated at the extremum $r_{0}$ of the function $Q_{0}$. In the eikonal limit for perturbations 
   \begin{equation}
   	Q_{0}\simeq \omega^2-f(r) \frac{l^2}{r^2},
   \end{equation}
   one can find that the extremum of $Q_{0}$ satisfies $2f(r_{0})-r_{0} f'(r_{0})=0$, which coincides with Eq. (\ref{eq:photon sphere orbit}) in the case of $\lambda=0$ i.e. the position of the effective potential’s extremum $r_{0}$ coincides with the location of the unmodified unstable photon orbit $r_{ps}$. Then, there is a main result drawn by Cardoso $et$ $al$. \cite{cardoso2009geodesic}. In the eikonal approximation, for any spherically symmetric spacetimes, the real part of the QNMs is given by the angular velocity of the unstable circular null geodesics while the imaginary part of the QNMs is related to the instability time scale of the orbit that can be represented by the Lyapunov exponent $\eta$
   \begin{equation}
   	\omega=\Omega l-i(n+1/2)|\eta|.
   \end{equation}
   Here, $\Omega$ and $\eta$ are given by
   \begin{gather}
   	\Omega=\frac{\sqrt{f(r_{ps})}}{r_{ps}}, \\
   	\eta=\sqrt{\frac{f(r_{ps})(2f(r_{ps})-r^2_{ps} f''(r_{ps}))}{2r^2_{ps}}},
   \end{gather}
   with $r_{ps}$ is the unmodified photon sphere radius. What's more, Stefanov $et$ $al$. published a paper in 2010 \cite{stefanov2010connection}, which pointed out a connection between the black hole QNMs and the strong gravitation lensing in the strong field regime. In recent years, Jusufi found that in the eikonal limit the real part of QNMs is inversely proportional to the shadow radius \cite{jusufi2020quasinormal,jusufi2020connection,jusufi2020quasinormal1}
   
   \begin{equation}
   	\omega_R \propto \frac{1}{R_{sh}}.
   \end{equation}
   Further, the relationship between the real part of QNMs and the shadow radius can be expressed as 
   
   \begin{equation}\label{eq:relation}
   		\omega_R = \lim_{l \gg 1} \frac{l}{R_{sh}},
   \end{equation}
  which is precise in the eikonal limit having large values of $l$ under the case of the photons uncoupled to the Weyl tensor. By this relationship, one obtains that the unmodified shadow radius decreases with the increase of the GB coupling constant $\alpha$ as the real part of QNMs increases. This effect is indeed shown to be the case in Figs. \ref{shadow2} and \ref{QNMs}. Besides, from Table \ref{t0} and Table \ref{t1}, one can find the numerical results are consistent with Eq. (\ref{eq:relation}) approximately. It can be interpreted as the fact that a decrease of the black hole horizon leads to a decrease of the shadow radius and consequently  higher values for the oscillation frequencies. In addition, the Lyapunov exponent $\eta$ is shown for the 4D Einstein-Gauss-Bonnet black hole in Fig. \ref{the Lyapunov exponent}. One can find that the variation of the Lyapunov exponent $\eta$ with the GB coupling parameter $\alpha$ in Fig. \ref{the Lyapunov exponent} is consistent with that of $\omega_{I}$ with $\alpha$ in Fig. \ref{QNMs}

\section{CONCLUSION}\label{6}
  In this parper, we have studied the shadow of a 4D Einstein-Gauss-Bonnet black hole as the photons couple to Weyl tensor and pointed out a simple connection between the  shadow radius and the real part of QNMs. We found that the shadow of black hole depends not only on the GB coupling constant $\alpha$, but also on the coupling parameter $\lambda$ and the polarization of photon. For different polarizations of photon, the shadow radius $R_{sh}$ decreases as the GB coupling constant $\alpha$ increases. However, with a increase of the coupling parameter $\lambda$, the shadow radius $R_{sh}$ increases for PPL and decreases for PPN. 
  
  Moreover, the connection between the shadow radius and the real part of QNMs satisfies Eq. (\ref{eq:relation}) while taking $\lambda=0$ in the spacetime background of a 4D Einstein-Gauss-Bonnet black hole. Computing the QNM frequencies for the scalar field perturbation by the sixth-order WKB approach, we found that the real part of QNMs increases with the increase of the GB coupling constant $\alpha$. Then using the correspondence between the geodesics and quasinormal spectrum, we pointed out the relationship between the shadow radius and the real part of QNMs under the case of the photons uncoupled to the Weyl tensor. We have also applied the unmodified unstable photon orbit of the black hole to compute the imaginary part of the QNM frequencies in the eikonal limit. This may imply a potential connection between the shadow of black hole and gravitational wave. Recently, the result that the size of the shadow can reflect the phase structure of the axially symmetric black hole was explored in Ref. \cite{zhang2020can}. This suggests that QNMs in connection with the thermodynamic phase structure of the black hole and our work opens an avenue for exploring this connection.

\section*{Conflicts of Interest}
The authors declare that there are no conflicts of interest regarding the publication of this paper.

\section*{Acknowledgments}
We would like to thank the National Natural Science Foundation of China (Grant No.11571342) for supporting us on this work.

\section*{References}

\bibliographystyle{unsrt}
\bibliography{ref}

\end{document}